\documentclass[preprint,showpacs,preprintnumbers,amsmath,amssymb]{revtex4}

\usepackage{./axodraw}
\usepackage{epsfig,latexsym,cancel,amssymb,amsmath}
\usepackage{color}
\usepackage{graphicx}

\newcommand{\be}{\begin{equation}}
\newcommand{\ee}{\end{equation}}
\newcommand{\bea}{\begin{eqnarray}}
\newcommand{\eea}{\end{eqnarray}}
\newcommand{\ba}{\begin{array}}
\newcommand{\ea}{\end{array}}

\newcommand{\Lint}{{\mathcal{L}_{\textrm{int}}}}

\begin{document}

\preprint{PI/UAN-2009-348FT}

\title{Color Octet Leptogenesis}

\author{Marta Losada}
\affiliation{Centro de Investigaciones, Cra 3 Este No 47A-15, Universidad Antonio Nari\~{n}o, Bogota, Colombia}
\author{Sean Tulin} 
\affiliation{California Institute of Technology, MC 452-48, 1200 California Blvd, Pasadena, CA 91125}

\date{\today}

\begin{abstract}

We study the implications for generating the cosmological baryon asymmetry through leptogenesis in the recent model of Fileviez Perez and Wise, which provides a new mechanism for generating neutrino masses at one-loop by introducing new color octet fermion and scalar fields.  We find that there are significant differences with respect to
other models for leptogenesis: low scale leptogenesis can occur naturally and the CP asymmetry can be large as there is no upper bound arising from neutrino masses. The CP asymmetry is insensitive to the phases in the neutrino mixing matrix. We investigate in detail the minimal model that can simultaneously fit low scale neutrino physics, the $\mu \rightarrow e \gamma$ bound and leptogenesis. The model can provide 
outstanding collider signatures and the value of the CP-asymmetry can be more constrained from lepton flavour violating processes than from neutrino physics.

\end{abstract}

\pacs{}
\maketitle

\section{Introduction}

Although it is well established that neutrinos have mass, the nature of their mass is still a mystery, requiring new degrees of freedom beyond the Standard Model (SM).  It is well known that if these degrees of freedom violate lepton number --- such that neutrinos are Majorana fermions --- then their dynamics during the early universe can satisfy the three Sakharov conditions \cite{Sakharov:1967dj} for the generation of the cosmological baryon asymmetry, via leptogenesis.

In leptogenesis, a lepton asymmetry is generated by the CP-violating, out-of-equilibrium decays of a heavy species $X$ \cite{fu86,review}. Assuming this occurs before the electroweak phase transition, this lepton asymmetry is partially converted into a baryon asymmetry by electroweak sphaleron transitions \cite{Kuzmin:1985mm}.  
The lepton-number-density-to-entropy-density ratio is parametrized in the following way:
\be
\frac{n_L}{s} = \epsilon_X \, \eta \, \left( \frac{n_X}{s} \right)_{T\gg m_X} \; .
\ee
The CP-violating asymmetry $\epsilon_X$ gives the average lepton number produced per $X$ decay.  In general, there can be several decaying species, each with their own CP-asymmetry.  The efficiency factor $\eta$ describes the competition between CP-violating decays and processes that washout lepton-number; it results from a full numerical solution of the Boltzmann equations.  In turn, the baryon asymmetry is $n_B/s = \kappa \, (n_L/s)$, with lepton-to-baryon conversion factor $\kappa = -28/51$ when only SM degrees of freedom are relativistic during sphaleron freeze-out.

Unfortunately, many leptogenesis scenarios involve energy scales far beyond the reach of colliders and are difficult to probe experimentally.  For example, consider a type-I see-saw scenario with hierarchical right-handed neutrinos, in which the lepton asymmetry is generated by the decays of the lightest, massive right-handed neutrino $N_1$.  In this case, the CP-asymmetry satisfies the bound \cite{Davidson:2002qv,Hamaguchi:2001gw,flavour}
\be
|\epsilon_{N_1}| \; \le \; \frac{3}{16\pi} \frac{m_{N_1}}{v^2} \,  m_{\nu_3}\;. \label{eq:bound}
\ee
The magnitude of $n_L$ is inextricably tied to the smallness of the neutrino masses; in this scenario a strong lower bound on the mass of the decaying right handed neutrino exists: $m_{N_1} \gtrsim 10^{9}$ GeV \cite{Davidson:2002qv}.  Furthermore, in the flavor blind case, the CP-violating phases that give rise to $\epsilon_{N_1}$ are not determined by the CP-violating phases that are in principle observable in the light neutrino sector. Only imposing certain assumptions on the high energy Lagrangian parameters
is it possible to relate leptogenesis quantities with low energy observables including rare decays; see \cite{Ibarra:2009bg} and references therein.
An alternative approach has been to identify possible scenarios for leptogenesis at a low energy scale \cite{Hambye:2001eu} through different mechanisms such as: mass degeneracy leading to resonant leptogenesis \cite{Pilaftsis:1997jf}, hierarchy in couplings, three body decays and the introduction of additional scales.

Recently, Fileviez Perez and Wise (FW) discovered several viable possibilities for generating Majorana neutrino masses at one-loop with new color octet scalar and fermion degrees of freedom~\cite{FileviezPerez:2009ud}.  The main attraction of these scenarios is that these color octets may be (but are not required to be) at the electroweak scale, thus accessible to the LHC.

In this work, we study leptogenesis in the FW model.  Here, the CP-violating decays are those of the octet fermions $F$ or scalars $S$.  As opposed to Eq.~\eqref{eq:bound}, the CP-asymmetries in the FW model are not constrained by the smallness of the light neutrino masses.  In principle, leptogenesis is possible with electroweak-scale masses for $F$ and $S$.  We note that in this model the enhancement of the CP-asymmetry does not rely on a hierarchy amongst the couplings constants that provide the necessary CP violating phases, nor mass degeneracies or three body decays.

We consider in detail the simplest leptogenesis scenario in the FW model: where the CP-asymmetry $\epsilon_{F_1}$ is driven by two-body decays of the lightest octet fermion $F_1 \to S \ell$.  Although $\epsilon_{F_1}$ is not constrained by neutrino masses, experimental limits on lepton flavor violating processes provide stringent bounds on the magnitude of $\epsilon_{F_1}$.  We compute the implications for $\mu \to e \, \gamma$.  We discuss how this simplest scenario can be extended in order to evade these constraints.

A significant difference with respect to the standard type I see-saw model of leptogenesis is the existence of the additional gauge interactions of the heavy decaying particle. It has been shown for different scenarios that the gauge interactions of the heavy decaying particle need not dilute the lepton asymmetry excessively \cite{Plumacher:1996kc,Racker:2008hp, Hambye:2003rt, Hambye:2005tk} both in the weak and strong washout regimes.
In our case the strong gauge interactions of the new color octet scalar and fermion degrees of freedom will permit them to
easily obtain a thermal abundance, strongly reducing the dependence on initial conditions\cite{Plumacher:1996kc}
.  Although the color octets will be kept closely in thermal equilibrium through gauge interactions, this does not necessarily preclude the generation of a significant lepton asymmetry if the decay rate is larger than the gauge annihilation rate.   Studies of  fermionic  and scalar SU(2)$_L$ triplet leptogenesis found that a sizable lepton asymmetry could be generated, with $\eta \sim \mathcal{O}(1)$, despite small departures from thermal equilibrium due to electroweak gauge interactions~\cite{Hambye:2003rt, Hambye:2005tk}.  We defer to future work a computation of the efficiency factor $\eta$ for leptogenesis in the FW model.  However, these strong gauge interactions are certainly beneficial from a phenomenological perspective in that they provide a mechanism to produce the color octets at colliders.  Indeed, this may be the most experimentally accessible leptogenesis scenario proposed so far.

The main purpose of this paper is to show that low scale leptogenesis in the FW model is feasible and compatible with constraints from neutrino physics and lepton flavor violation.  In section II, we review the generation of neutrino masses in the FW model.   Section III
provides the main results for the CP-asymmetry relevant for leptogenesis and the
correlation with neutrino physics.  In Section IV, we identify the constraints from lepton flavor violating processes and briefly comment on collider signatures. In the final section we summarize the main results.

\section{Neutrino masses from color octets}

\label{sec:model}

In the FW model, the Standard Model is extended through the inclusion of the following additional fields: (i) $\mathcal{N}_S$ scalar fields $S$ with SU(3)$_C\times$SU(2)$_L\times$U(1)$_Y$ quantum numbers (8,2,1/2), and (ii) $\mathcal{N}_F$ fermions $F$ with quantum numbers (8,1,0) \footnote{For simplicity, we focus on only one choice for the SU(2)$_L\times$U(1)$_Y$ quantum numbers of $S$ and $F$. Other options are also viable \cite{FileviezPerez:2009ud}.}.  We couple these fields to the SM through the most general gauge invariant and renormalizable interaction Lagrangian (using two-component spinor notation)
\begin{widetext}
\be
\Lint = \left( \, y_{iab} \, L^i \, \epsilon \, F_a \, S_b + g_{ijb}^u \, u_{R}^{i\dagger} \, S_b \epsilon \, Q^j   + g_{ijb}^d \, d_{R}^{i\dagger} S_b^\dagger \, Q^j  + \textrm{h.c.} \, \right) \; - \; V(H,S) \; .
\ee
\end{widetext}
The scalar potential $V$ contains many terms \cite{Manohar:2006ga}; here, the only one of relevance is
\be
V \supset - \lambda_{bc} \, S_b^\dagger H \, S_c^\dagger H  + \; \textrm{h.c.} \; .
\ee
Our notation is as follows: the SU(2)$_L$ doublets are $L_i = (\nu_L^i, \, e_{L}^i)$, $Q_i = (u_{L}^i, \, d_{L}^i)$, $H=(H^+, \, H^0)$, and $S=(S^+, \, S^0)$; the indices $i,j=1,2,3$ label generation, while indices $a=1..\mathcal{N}_F$, $b,c=1..\mathcal{N}_S$ label the new fields; we have suppressed SU(3)$_C\times$SU(2)$_L$ indices; and lastly, the antisymmetric tensor $\epsilon$ (with $\epsilon_{12} = 1$) acts on the SU(2)$_L$ isospin space.  

This scenario provides a new mechanism for generating neutrino masses~\cite{FileviezPerez:2009ud}.  Following these authors, we assume that $\lambda$ is diagonal: $\lambda_{bc} \equiv \lambda_b \, \delta_{bc}$.  The left-handed neutrino Majorana mass matrix, arising at one-loop order, is
\be
M_{ij}^\nu = \sum_{ab} \, \frac{1}{4\pi^2} \: y_{iab} \, y_{jab} \, \lambda_b \, v^2 \, m_{F_a} \, \frac{m_{S_b}^2 + m_{F_a}^2 \left( \log(m_{F_a}^2/m_{S_b}^2) - 1 \right) }{\left( m_{F_a}^2 - m_{S_b}^2 \right)^2 } \; .
\ee
The eigenvalues and mixing angles of this matrix are constrained by neutrino oscillation experiments.  The minimum field content needed to reproduce these constraints is either $\mathcal{N}_S = 1$ and $\mathcal{N}_F = 2$, or $\mathcal{N}_S = 2$ and $\mathcal{N}_F = 1$~\cite{FileviezPerez:2009ud}; in these cases, the lightest neutrino is massless.  This scenario can accomodate three massive neutrinos for $\mathcal{N}_S = \mathcal{N}_F = 2$.  

In our analysis below, we consider both the normal and inverted hierarchies, with the following input parameters for the light neutrino masses:
\be
(m_{\nu_1}, \, m_{\nu_2}, \, m_{\nu_3} ) = \left\{ \ba{lll} \left( 0, \, \sqrt{\Delta m^2_{\textrm{sol}}} , \, \sqrt{\Delta m^2_{\textrm{atm}}} \right) & \qquad & \textrm{[normal]} \\
\left(  \sqrt{\Delta m^2_{\textrm{atm}} - \Delta m^2_{\textrm{sol}}} , \, \sqrt{\Delta m^2_{\textrm{atm}}} , \, 0 \right) & \qquad & \textrm{[inverted]} \ea \right.
\ee
where $\Delta m^2_{\textrm{sol}}\simeq 8\times 10^{-5}$ eV$^2$ and $\Delta m^2_{\textrm{atm}}\simeq 8\times 10^{-3}$ eV$^2$, 
and mixing angles $\theta_{12} \simeq 35^\circ$, $\theta_{23} \simeq 45^\circ$, and $\theta_{13} \simeq 0$~\cite{Amsler:2008zzb}.  We neglect potential Majorana phases in the neutrino mixing matrix.

\section{Leptogenesis}

\label{sec:lepto}

Within the FW model, there are several different leptogenesis scenarios, depending on the number of octet fermions and scalars, and the hierarchy of their masses.  In this section we consider the simplest case, with $\mathcal{N}_S = 1$ and $\mathcal{N}_F = 2$, and compute the relevant CP-violating asymmetry.     
The overall scale of the light neutrino masses does not constrain the CP-violating asymmetries.  As we show below,
the decay rate and the asymmetry are proportional to the coupling constants $y$, but do not involve $\lambda$.  As far as neutrino masses are concerned, we can have $y \sim \mathcal{O}(1)$, with TeV-scale masses $m_F$ and $m_S$, as long as we tune $\lambda$ to be sufficiently small: $\lambda \lesssim 10^{-10}$.

\begin{figure}[!t]
\begin{center}
\mbox{\hspace*{-0.7cm}\epsfig{file=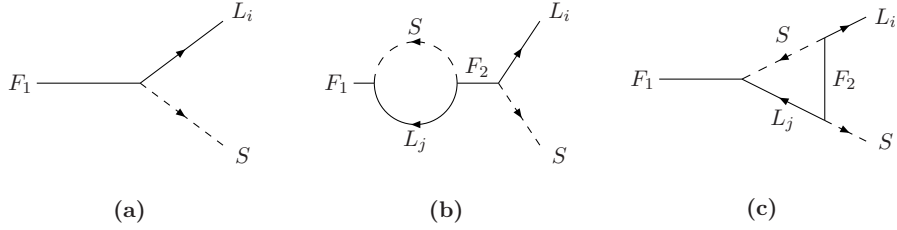,height=3cm}}
\end{center}
\caption{\it \small 
Tree-level (a) and one-loop (b,c) amplitudes for $F_1 \to S L_i$ decay.  The interference between these amplitudes gives rise to the CP-asymmetry $\epsilon_{F_1}$.  
}
\label{fig:decay}
\end{figure}

Let us consider the case with masses $m_{F_2} > m_{F_1} > m_{S}$.  We assume that the CP-asymmetry is driven primarily by decays of $F_1$, shown in Fig.~\ref{fig:decay}.  This assumption is most valid when $m_{F_2} \gg m_{F_1}$, so that $F_2$ freezes out while $F_1$ is in equilibrium; any CP-asymmetry generated by the decays of $F_2$ will be washed out by processes $L S \leftrightarrow F_1 \leftrightarrow S^\dagger \bar{L}$.

The tree-level decay rate is
\be
\left. \frac{}{} \Gamma(F_1 \to S L_i) \right|_{\textrm tree} = \left. \frac{}{} \Gamma(F_1 \to S^\dagger \bar L_i) \right|_{\textrm tree}= \frac{|y_{i1}|^2}{16\pi} \, \frac{(m_{F_1}^2 - m_S^2)^2}{m_{F_1}^3} \; .
\ee
Here, we have suppressed the label $b$ for $S_b$, since $b=1$ only.

The CP-violating asymmetry, defined as
\be
\epsilon_{F_1} \equiv \frac{\sum_i \Gamma(F_1 \to S L_i) - \Gamma(F_1 \to S^\dagger \bar{L}_i) }{\sum_i \Gamma(F_1 \to S L_i) + \Gamma(F_1 \to S^\dagger \bar{L}_i) }  \; ,
\ee
is non-zero due to the interference between tree-level (Fig.~\ref{fig:decay}a) and one-loop amplitudes (Fig.~\ref{fig:decay}b,c).  We find the asymmetry
\be
\epsilon_{F_1} = \frac{3}{8\pi} \, \frac{\sum_{i,j} \textrm{Im}[y_{i1} y_{i2}^* y_{j1} y_{j2}^*]}{\sum_i |y_{i1}|^2} \, \frac{(m_{F_1}^2 - m_S^2)^2}{m_{F_1}^3 m_{F_2}} \, f(m_{F_1}, \, m_{F_2}, \, m_S) \; ,\label{CP}
\ee
where we have defined
\begin{align}
f(m_{F_1}, \, m_{F_2}, \, m_S) &\equiv \frac{2 \, m_{F_2}^2}{3 (m_{F_2}^2 - m_{F_1}^2)(m_{F_1}^2 - m_{S}^2)^4} \\
\times & \; \left\{ \frac{}{} \right. m_{F_1}^4 (m_{F_2}^2 - m_{F_1}^2)(m_{F_1}^2 + m_{F_2}^2- 2 m_S^2) \log \left[ \frac{m_{F_1}^2 (m_{F_1}^2 + m_{F_2}^2- 2 m_S^2)}{m_{F_1}^2 \, m_{F_2}^2 - m_S^4 } \right] \notag \\
&\; \qquad \left. \frac{}{} + (m_{F_1}^2 - m_S^2)^2 (2 m_{F_1}^4 + m_S^4 - m_{F_1}^2 m_{F_2}^2 - 2 \, m_{F_1}^2 m_S^2) \right\} \notag
\end{align}
such that $f(m_{F_1},m_{F_2},m_S) = 1$ in the limit that $m_{F_2} \gg m_{F_1}$.

We now show that the CP-violating phases that enter into the light-neutrino mixing matrix are {\it not} relevant to the CP-asymmetry that drives leptogenesis. 
When $\mathcal{N_S} = 1$ and $\mathcal{N}_F = 2$, the neutrino mass matrix has the form
\be
M_{ij}^\nu = \sum_a \, M_a \, y_{ia} \, y_{ja} \; ,
\ee
where
\be
M_a \equiv  \frac{\lambda \, v^2 \, m_{F_a}}{4\pi^2}  \, \frac{m_{S}^2 + m_{F_a}^2 \left( \log(m_{F_a}^2/m_{S}^2) - 1 \right) }{\left( m_{F_a}^2 - m_{S}^2 \right)^2 } \; ,
\ee
and we have suppressed the $b$ index, since $b=1$ only.  In this case, the most general form for $y$ that gives the correct light neutrino masses and mixing angles is $y = U \cdot X$, where for the normal hierarchy
\be
X \equiv \left( \ba{cc} 0 & 0 \\ \eta_1 \, \sqrt{ \frac{m_{\nu_2}}{M_1} - \frac{m_{\nu_2}}{m_{\nu_3}} \, x^2} & \eta_2 \, \sqrt{ \frac{M_1 \, m_{\nu_2}}{M_2 \, m_{\nu_3}}  } \, x \\ x & - \eta_1 \eta_2 \, \sqrt{ \frac{m_{\nu_3}}{M_2} - \frac{M_1}{M_2} \, x^2} \ea \right) \; , \label{eq:decomp}
\ee
and for the inverted hierarchy
\be
X \equiv \left( \ba{cc} \eta_1 \, \sqrt{ \frac{m_{\nu_1}}{M_1} - \frac{m_{\nu_1}}{m_{\nu_2}} \, x^2 } & \eta_2 \, \sqrt{ \frac{M_1 \, m_{\nu_1}}{M_2 \, m_{\nu_2}}  } \, x \\ x & - \eta_1 \eta_2 \, \sqrt{ \frac{m_{\nu_2}}{M_2} - \frac{M_1}{M_2} \, x^2}  \\ 0 & 0 \ea \right) \; . \label{eq:decomp2}
\ee
Here, $x$ is an undetermined complex parameter, and $\eta_i = \pm 1$.  The matrix $U$ is the neutrino mixing matrix, containing the mixing angles and phases that are in principle observable through studies of light neutrinos.  
The CP-asymmetry is proportional to the factor
\be
\sum_{i,j} \textrm{Im}[y_{i1} y_{i2}^* y_{j1} y_{j2}^*] = \sum_{i,j} \textrm{Im}[X_{i1} X_{i2}^* X_{j1} X_{j2}^*] \; ,
\ee
where the right side follows by the unitarity of $U$.  Therefore, we find that the CP-violating phases that drive leptogenesis are contained in $X$ and are independent of the phases in $U$.  It is not difficult to generalize this argument to any $\mathcal{N}_S$ and $\mathcal{N}_F$.

Some additional comments are now in order:

\begin{itemize}
 
\item For a full calculation in the context of our model for leptogenesis at low temperatures flavor effects \cite{flavour} must be taken into account.  There would be an additional diagram that contributes to each CP flavour asymmetry which corresponds to the self energy correction with the lepton flowing in the bubble in the opposite direction.
In this case  the CP-asymmetry will also depend on the the $U$ matrix elements. However, it has been shown
\cite{AristizabalSierra:2009mq} that when rates for lepton flavour violating processes are large at high temperature the chemical potentials for leptons will equilibrate and thus the CP-asymmetry is correctly calculated by eq.(\ref{CP}).
For large portions of the region of parameters that we study the  scatterings involving lepton flavour violation (LFV) are in equilibrium.
See below for the implications of LFV in low energy processes.

\item For the regime in which $m_{S_{1}} > m_{F_{i}} + m_l$ the CP asymmetry will be produced by the decay of the scalar octets.
In this case the minimum particle content required is of two scalar and one fermion octet to be consistent with neutrino physics. The expression for the CP-asymmetry can be similarly obtained.

\item For some specific mass regimes, such as $m_{F_{2}}> m_{S_{1}} >m_{F_{1}}$, the CP asymmetry can be produced through three body decays.

\end{itemize}

\begin{figure}[!t]
\begin{center}
\mbox{\hspace*{-0.7cm}\epsfig{file=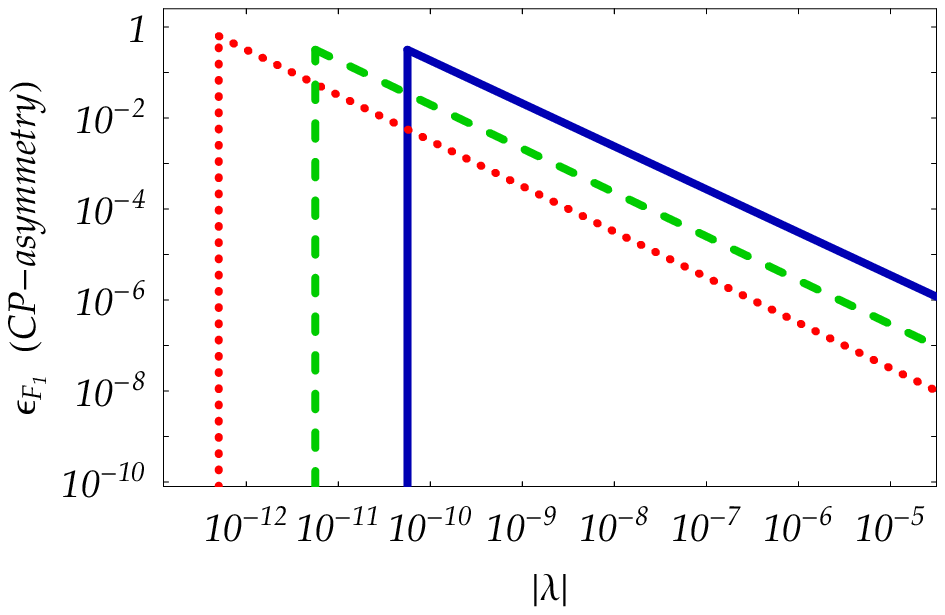,height=5cm}} \mbox{\hspace*{0cm}\epsfig{file=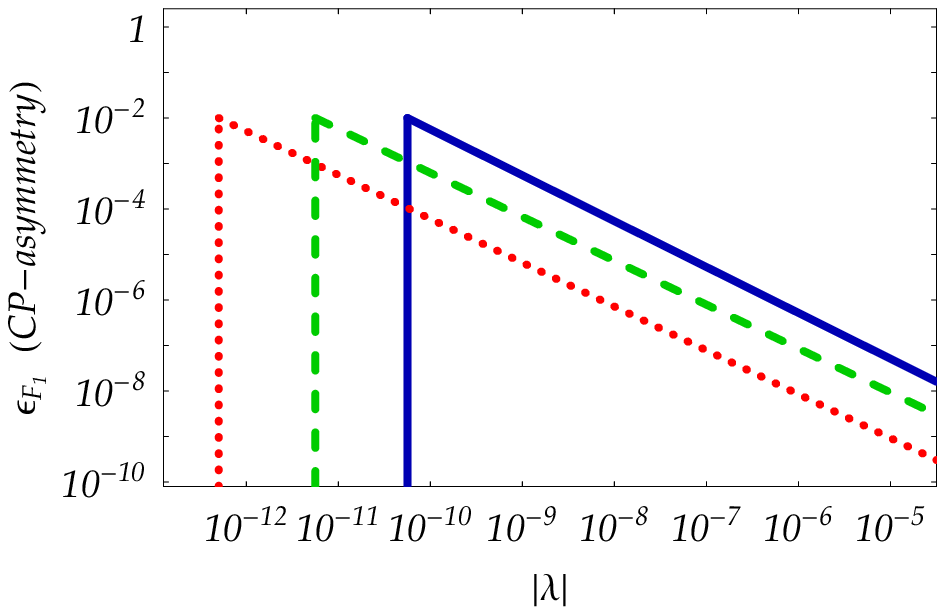,height=5cm}}
\end{center}
\caption{\it\small 
Results of our parameter scan, showing the correlation between the CP-asymmetry $\epsilon_{F_1}$ and the $(S^\dagger H)^2$ coupling $\lambda$. 
Areas under curves denote regions of viable parameter space, consistent with perturbativity in $y$ and with observed light neutrino masses and mixing angles for normal (left panel) and inverted (right panel) hierarchies.  The multiple curves denote different choices of octet masses: $(m_S, \, m_{F_1}, \, m_{F_2}) = (0.1, \, 0.2, \, 0.5) \; \textrm{TeV}$ (dotted red), $(1, \, 2, \, 5) \; \textrm{TeV}$ (dashed green), and $(10, \, 20, \, 50) \; \textrm{TeV}$ (solid blue).}
\label{fig:scan}
\end{figure}

We now compute the (unflavored) CP-asymmetry numerically.  Our results, described below, are shown in Fig.~\ref{fig:scan}.  We take three representative choices for the scalar and fermion octet masses: 
\be
(m_S, \, m_{F_1}, \, m_{F_2}) = \left\{ \ba{lll} (0.1, \, 0.2, \, 0.5) & \textrm{TeV} & \textrm{[dotted red]} \\ (1, \, 2, \, 5) & \textrm{TeV} & \textrm{[dashed green]} \\ (10, \, 20, \, 50) & \textrm{TeV} & \textrm{[solid blue]} \ea \right.
\ee
Next, we perform a parameter scan over $\lambda$ and $x$; according to Eqs.~(\ref{eq:decomp},\ref{eq:decomp2}), these are the only remaining free parameters in $y$ if we enforce consistency with the observed light neutrino mass parameters combined with the normal or inverted hierarchy cases.  We scan over the following range:
\be
\label{eq:masses}
10^{-14} < |\lambda| < 10^{-4} \; , \qquad 10^{-6} < |x| < 10 \; , \qquad 0 < \arg(x) < 2\pi \; .
\ee
Furthermore, we randomly choose the signs of $\lambda$ and $\eta_{1,2}$.  We impose perturbativity in $y$ by discarding parameter points for which $\textrm{max}(y_{ia}) > \sqrt{4\pi}$.  The upper bound on $|x|$ is essentially arbitrary and has been chosen so that it is possible to saturate the perturbativity bound on $y$ during our scan.  On the other hand, in the limit that $|x| \to 0$, the matrix $X$ is real and the CP-violating asymmetry vanishes.

In Fig.~\ref{fig:scan}, we show regions of parameter space consistent with our numerical scans.  In both panels, we plot the CP-asymmetry $\epsilon_{F_1}$ as a function of $|\lambda|$.  In the left panel, the area under the red dotted curve shows the region of viable parameter space given by our scan for octet masses $(m_S, \, m_{F_1}, \, m_{F_2}) = (0.1, \, 0.2, \, 0.5) \; \textrm{TeV}$, consistent with all observed neutrino masses and mixing angles for the normal hierarchy.  The upper, diagonal bound arises due to the neutrino masses, while the left bound arises due to perturbativity in $y$.  The other curves show the corresponding regions of viable parameter space for heavier choices of octet masses: $(m_S, \, m_{F_1}, \, m_{F_2}) = (1, \, 2, \, 5) \; \textrm{TeV}$ (dashed green) and $(m_S, \, m_{F_1}, \, m_{F_2}) = (10, \, 20, \, 50) \; \textrm{TeV}$ (solid blue).  In the right panel, we show the same results for the inverted hierarchy case.  Compared to the normal case, the CP-asymmetry in the inverted case tends to be suppressed by two orders of magnitude; numerically, this occurs due to the near-degeneracy of $m_{\nu_1} \simeq m_{\nu_2}$.

Our main conclusion is that the CP-asymmetry can be as large as $\mathcal{O}(1)$ in this model for normal hierarchy and $\mathcal{O}(10^{-2})$ for the inverted hierarchy of the light neutrino masses.  Assuming that washout processes are not severe, this appears promising for leptogenesis.  For fixed $|\lambda|$, the CP-asymmetry can be increased for larger octet masses; on the other hand, for fixed octet masses, the CP-asymmetry can be increased by decreasing $|\lambda|$.

\section{Lepton flavor violation}

We have seen that in this model for leptogenesis the scale of the masses of the new degrees of freedom can be on the order of
the electroweak scale, with sizeable couplings to leptons.  This opens up the possibility of
new signatures at colliders and in low energy precision measurements.

In this work, we focus on an important constraint from experimental limits on the lepton flavor violating decay $\mu \rightarrow e \gamma$.  Although the rate for this process is independent of the CP-violating phases relevant for leptogenesis, it
can strongly constrain the magnitude of some of the couplings that enter the CP-asymmetry.  For arbitrary $\mathcal{N}_F$ and $\mathcal{N}_S$, the branching ratio is
\be
\textrm{Br}[\mu \rightarrow e \gamma] = \frac{12\pi^2 \, |A|^2}{G_F^2 \, m_\mu^2} \; ,
\ee
where
\be
\label{eq:amp}
A \equiv \sum_{a,b} \frac{e \, y_{1ab}^* \, y_{2ab}}{32 \, \pi^2} \, m_\mu \, \int^1_0 dx \, \frac{1-x+x^2-x^3}{x\, m_{F_a}^2 + (1-x) \, m_{S_b}^2} \; .
\ee

\begin{figure}[!t]
\begin{center}
\mbox{\hspace*{-0.5cm}\epsfig{file=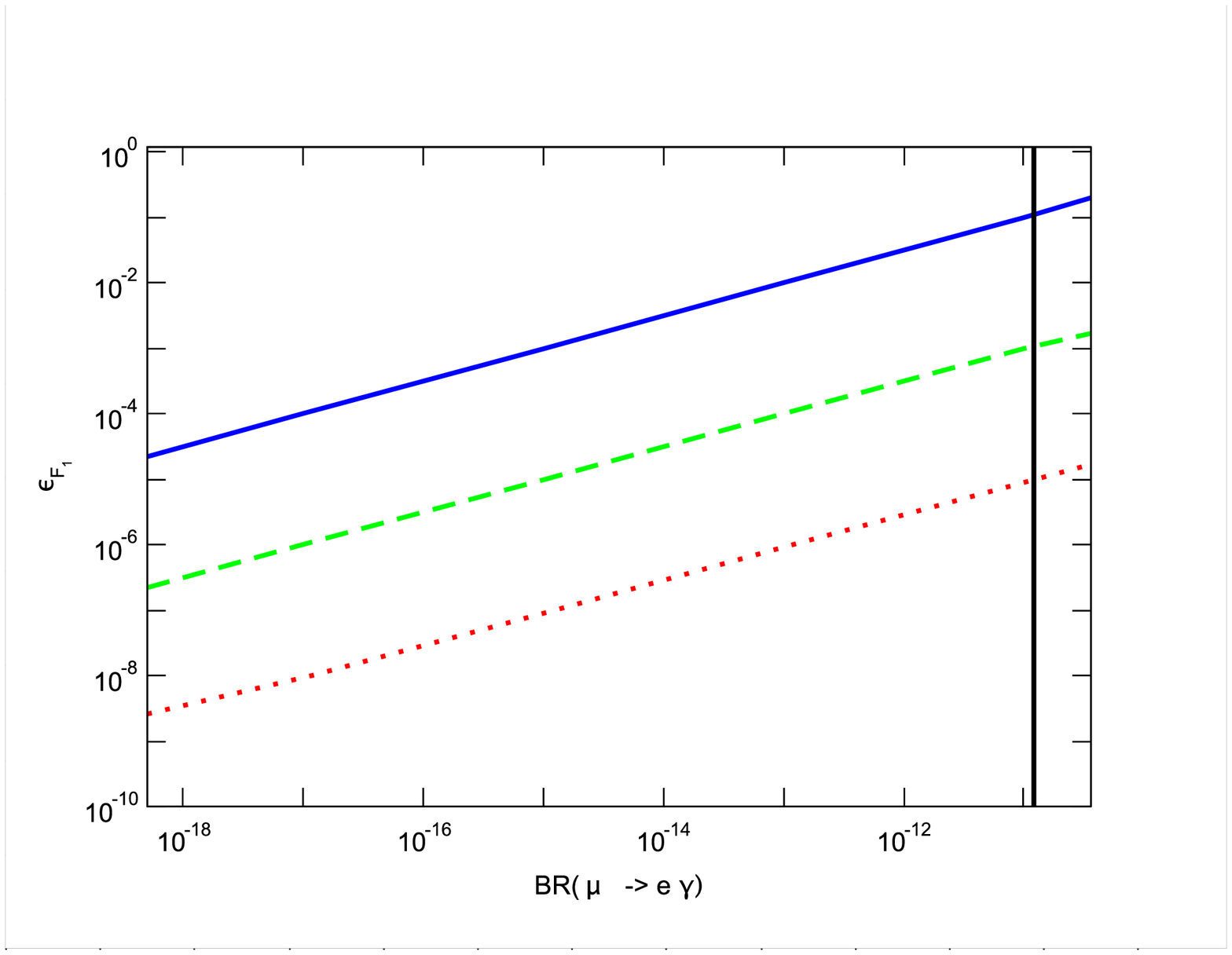,height=3.5cm}} \mbox{\hspace*{0cm}\epsfig{file=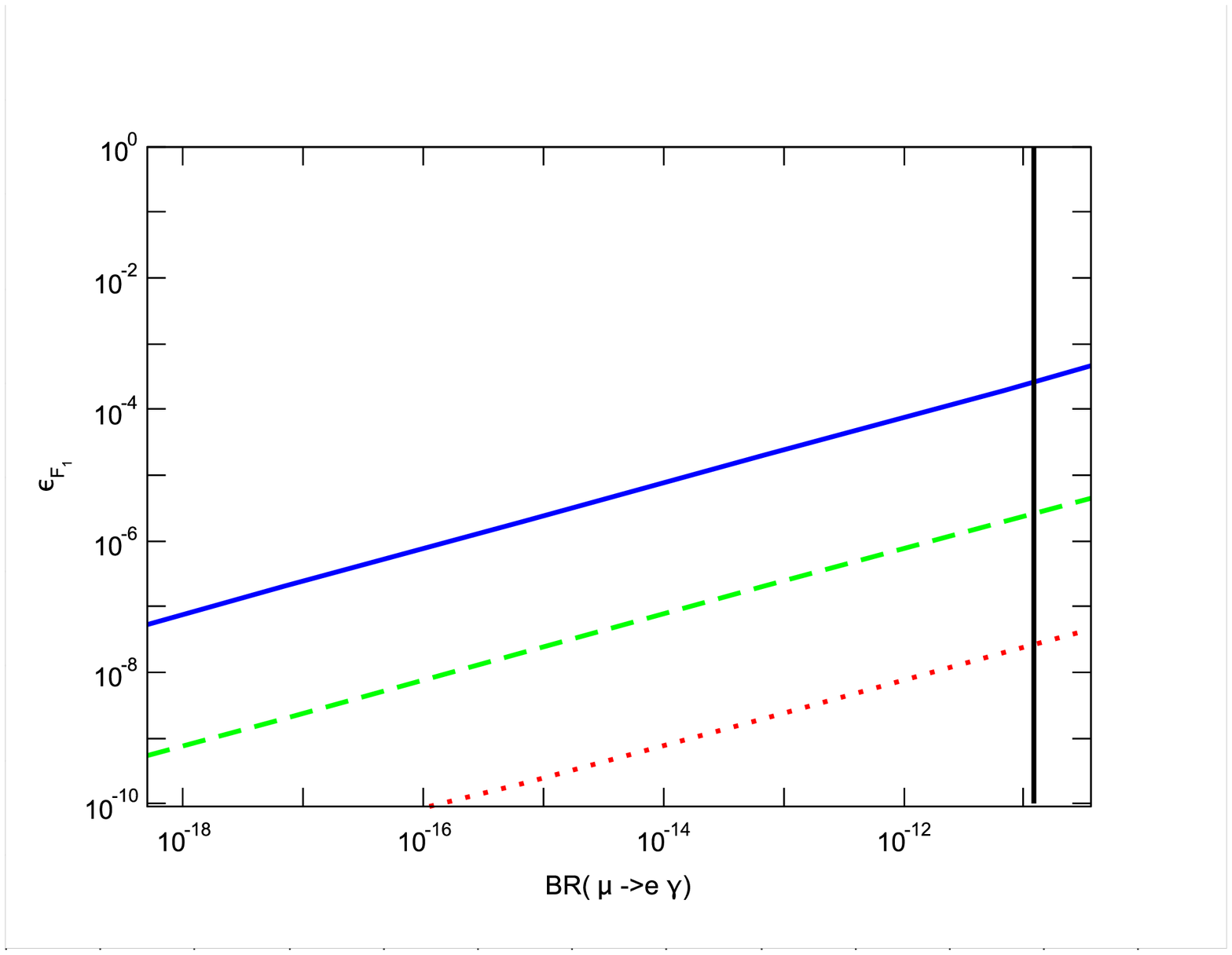,height=3.5cm}}
\end{center}
\caption{\it\small 
Parameter scan showing correlation between CP-asymmetry $\epsilon_{F_1}$ and $\textrm{Br}(\mu \rightarrow e\gamma)$ for three set of values of $m_{F_2}$, $m_{F_1}$ and $m_S$, as in Fig.~\ref{fig:scan}, for normal (left panel) and inverted (right panel) hierarchies.  Region under the curve is allowed parameter space consistent with neutrino masses and mixings.  Vertical line denotes present experimental limit on $\textrm{Br}(\mu \rightarrow e\gamma)$.
}
\label{fig:muegamma}
\end{figure}

Now, let us specialize to the case of our illustrative example, where $\mathcal{N}_S=1$ and $\mathcal{N}_F = 2$, and $m_{F_2} > m_{F_1} > m_S$.  In Fig.~\ref{fig:muegamma},
we show the results of our parameter scan, plotting the value of the CP-asymmetry versus the branching ratio of $\mu \rightarrow e \gamma$.   The regions under the curves correspond to viable parameter space regions for the normal (left panel) and inverted hierarchies (right panel), for three choices of octet masses listed in Eq.~\eqref{eq:masses}.  The vertical line denotes the current experimental bound $\textrm{Br}(\mu \rightarrow e\gamma) \le 1.2\times 10^{-11}$~\cite{Brooks:1999pu}, which strongly constrains the magnitude of the CP-asymmetry.  Limits on other lepton-flavor violating decays may provide additional important constraints; we defer a systematic analysis to a future study.  

There are two ways to evade these severe constraints.  First, one can increase the scale of the octet masses.  This is clearly shown in Fig.~\ref{fig:muegamma}, where for larger octet masses, the region of viable parameter space includes larger values of $\epsilon_{F_1}$.  Furthermore, it is sufficient to increase only $m_{F_{1,2}}$; the scalar can still be light.  

Second, one could consider an octet sector extended beyond the simplest case considered here.  With more parameters, there is more freedom to have large CP-violating asymmetries with supressed lepton flavor violation.  As a proof of principle, let us consider an example in which $\mathcal{N}_F = 2$ and $\mathcal{N}_S = 3$.  Next, suppose that the $y$ couplings could be written as $y_{iab} \equiv y_a \, \delta_{ib}$.  In this case, the $(L_i F_a S_b)$-vertex by itself will not lead to lepton flavor violating decays such as $\mu \to e \gamma$; according to Eq.~\eqref{eq:amp}, we have $A = 0$.  Meanwhile, suppose leptogenesis is driven by decays $F_1 \to L_i S_b$, similar to the previous section.  The CP-violating asymmetry will be proportional to
\be
\epsilon_{F_1} \; \propto \; \sum_{i,j} \sum_{b,c} \textrm{Im}[ y_{i1b} y_{j1b} y_{i2b}^* y_{j2a}^* ] = \textrm{Im}\left[ (y_1 y_2^*)^2 \right]
\ee
which can be large and non-zero for $\sin[\textrm{arg}(y_1 y_2^*)] \ne 0$.  On the other hand, lepton flavor violation must appear in the neutrino mass matrix; this can be induced with a non-diagonal parameter $\lambda_{bc}$.  If this parameter is sufficiently small, lepton flavor violation will be suppressed, while the CP-asymmetry will be enhanced, analogous to Fig.~\ref{fig:scan}.  This example shows that although lepton flavor violation constraints are strong, they can be evaded without pushing the octet masses beyond the electroweak scale.

It is also important to note that there is no significant constraint arising at one loop level from electric dipole moments
in the minimal model we have considered as the CP violating phases cancel out.

At the LHC the production of the fermion octet could occur similarly to the pair production of gluinos and the decay of the fermion octet with leptons in the final state would  produce  a distinctive signature of same-sign dileptons 
\cite{FileviezPerez:2009ud}. Also the constraints from electroweak precision constraints have been studied in detail
from the presence of scalar octets in \cite{Burgess:2009wm}, focusing on the so-called oblique corrections to the gauge boson vacuum polarizations.
Here we note that it is conceivable that additional constraints arising from electroweak precision observables can arise  on the models we have focused on from
corrections of the $Z\ell \ell$ vertex through $F$ and $S$ loops. 
However, a detailed analysis of these effects is beyond the scope of this paper.

\section{Summary}

We have explored the possibility of generating the cosmological baryon asymmety in the context of an extension of the SM with additional scalar and fermion octets. We have shown that it is possible to have a large value of the CP-asymmetry relevant for leptogenesis  at an energy scale not much larger than the electroweak scale and simultaneously fulfill experimental results for neutrino masses and mixings. The CP-asymmetry does not depend on the phases in the neutrino mixing matrix. We have not performed a detailed calculation of the efficiency of the production of the lepton asymmetry but note that it may be possible to
have cases in which the dilution from gauge interactions is not too significant. The fact that low scale leptogenesis can occur via the decay of a particle with SU(3) gauge interactions enhances its direct testability in the near future.
We have also indicated the relevance that laboratory bounds from the non-observance of rare leptonic decays can have on the magnitude of the CP-asymmetry. In particular, this may require that the color octets be on the order of a few TeV.

\begin{acknowledgments}

We thank Daniel J. H. Chung and Mark Wise for insightful discussions.  ML thanks Sacha Davidson for useful comments and suggestions and CCPP for hospitality during
the completion of this work.

\end{acknowledgments}

\end{document}